\documentclass[printer]{aa}  

\usepackage{graphicx}
\usepackage{txfonts}
\usepackage[colorlinks=true, allcolors=blue]{hyperref}

\usepackage{color}

\defcitealias{GPTA}{GPTA Paper I}
\bibpunct{(}{)}{;}{a}{}{,} 

\begin{document}

   \title{Timing Gamma-ray Pulsars using Gibbs Sampling}

   \author{C. J. Clark
     \inst{1,2}
     \and S. Valtolina
     \inst{1,2,3}
     \and L. Nieder
     \inst{1,2}
     \and R. van Haasteren
     \inst{1,2}
   }

   \institute{Max Planck Institute for Gravitational Physics (Albert Einstein Institute), D-30167 Hannover, Germany \\
     \email{colin.clark@aei.mpg.de}
     \and
     Leibniz Universit\"{a}t Hannover, D-30167 Hannover, Germany
     \and
     Max-Planck-Institut f\"{u}r Radioastronomie, Auf dem H\"{u}gel 69, D-53121 Bonn, Germany
}

   \date{Received XXX; accepted YYY}

 
   \abstract{ Timing analyses of gamma-ray pulsars in the \textit{Fermi}
     Large Area Telescope data set can provide sensitive probes of many
     astrophysical processes, including timing noise in young pulsars, orbital
     period variations in redback binaries, and the stochastic gravitational wave
     background (GWB). These goals can require careful accounting of stochastic
     noise processes, but existing methods developed to achieve this in
     radio pulsar timing analyses cannot be immediately applied to the discrete
     gamma-ray arrival time data. To address this, we have developed a new
     method for timing gamma-ray pulsars, in which the timing model fit is
     transformed into a weighted least squares problem by randomly assigning
     each photon to an individual Gaussian component of a template pulse
     profile. These random assignments are then numerically marginalised over
     through a Gibbs sampling scheme. This method allows for efficient
     estimation of timing and noise model parameters, while taking into account
     uncertainties in the pulse profile shape. We simulated \textit{Fermi}-LAT
     data sets for gamma-ray pulsars with power-law timing noise processes,
     showing that this method provides robust estimates of timing noise
     parameters. We also describe a Gaussian-process model for orbital period
     variations in black-widow and redback binary systems that can be fit using
     this new timing method. We demonstrate this method on the black-widow
     binary millisecond pulsar B1957$+$20, where the orbital period varies
     significantly over the LAT data, but which provides one of the most
     stringent gamma-ray upper limits on the GWB. }


   \maketitle
%

   \section{Introduction}
   \label{s:intro}
   The \textit{Fermi} Large Area Telescope \citep[LAT,][]{LAT} data set contains
   a near-continuous all-sky history of the arrival times of gamma-ray photons
   detected since \textit{Fermi}'s launch in 2008. Significant pulsations from
   more than 300 pulsars have been detected in the LAT data \citep{3PC}, split
   nearly evenly between energetic young pulsars and rapidly spinning recycled
   millisecond pulsars (MSPs).

   While pulsar timing studies, in which models for pulse arrival times are
   built to keep track of pulsar rotations, have long been applied to radio
   pulsar observations, a wealth of astrophysical processes can also be probed
   by timing gamma-ray pulsations in the LAT data. In fact, around 80 pulsars
   can only be reliably timed in gamma-ray data \citep{3PC} either because they
   appear to be radio quiet \citep[e.g.][]{Abdo2009+16PSRs,Pletsch2011}, or
   because their radio pulses are nearly always eclipsed by intra-binary
   material
   \citep{Ray2013+J1311,Ray2020+J2339,Nieder2020+J1653,Thongmeearkom2024+TRAPUMRBs}. Although
   a minor fraction of the total pulsar population, these pulsars represent some
   of the most energetic and most compact binaries, respectively.  Furthermore,
   the LAT data allows pulsars to be timed retroactively after their discovery,
   providing an immediate multi-year timing baseline, something that is only
   possible in radio data for a small number of special sky locations with
   multiple archival observations, e.g. globular clusters
   \citep{Balakrishnan2023+M30B,Padmanabh2024+Terzan5}.

   Characterisation of timing noise \citep[TN, stochastic variations in
     rotational frequency, e.g.][]{Kerr2015}, glitches \citep[sudden,
     instantaneous jumps in frequency,
     e.g.][]{Pletsch2012+J1838,Panin2025+J1522}, and correlated spin-down/flux
   variability \citep{Allafort2013+J2021,Takata2020+J2021, Fiori2024+J2021} in
   young gamma-ray pulsars can probe the dynamic pulsar magnetosphere and
   neutron star structure. Gamma-ray timing of binary MSPs has shown that
   complicated orbital period variations are nearly universal in redback
   binary systems \citep[e.g.][]{Pletsch2015+J2339,
     Thongmeearkom2024+TRAPUMRBs}, and has revealed a possible planetary-mass
   tertiary in one system \citep{Nieder2022+J1555}. The construction of a
   gamma-ray pulsar timing array (GPTA) can also provide an important
   cross-check of the searches for and characterisation of the stochastic
   gravitational wave background (GWB) from radio pulsar timing array (PTA)
   projects \citep[][hereafter \citetalias{GPTA}]{GPTA}. While the GPTA is less
   sensitive than radio PTAs, it is immune to variations in the interstellar
   medium that disperses and scatters radio pulsations, which can bias radio
   timing models.

   While sophisticated radio pulsar timing methods and models have been
   developed \citep[e.g.][]{TempoNest,enterprise,Susobhanan2025+Vela},
   particularly for PTA experiments, these are often not applicable to gamma-ray
   data, meaning dedicated gamma-ray timing methods have to be developed to make
   the most of this data set.  In the first gamma-ray timing studies
   \citep{Ray2011,Kerr2015}, several weeks or months of gamma-ray data were
   integrated to produce statistically significant integrated pulse
   profiles. These could then be compared to a template pulse to extract a small
   number of times-of-arrival (ToAs) that could be analysed using existing radio
   pulsar timing tools. However, many gamma-ray pulsars are faint enough that
   the required ToA integration time scales are longer than those of important
   timing effects such as binary Doppler shifts, or variations in the
   solar-system R\o{}mer delay, meaning binary and astrometry parameters must
   be assumed in advance, and cannot then be fit from the extracted ToAs. This
   procedure also assigns a symmetrical Gaussian uncertainty to each ToA, an
   assumption that may not be accurate when pulsation signal-to-noise is low.

   Later works \citep[e.g.][]{Pletsch2015+J2339,Clark2015+J1906} developed a
   fully unbinned approach that optimises the Poisson likelihood for the
   observed photon arrival times, given an assumed pulse profile shape. This
   technique was then further extended to also marginalise over the uncertain
   shape of the pulse profile \citep{Nieder2019+J0952}. This method uses the
   information from every individual photon arrival time, allowing all binary
   and astrometric parameters to be fit to the data. However, it requires a
   multi-dimensional MCMC sampling process that can become inefficient as the
   number of timing model parameters increases, and was not able to fit for
   stochastic noise components. In \citetalias{GPTA}, a similar unbinned
   photon-by-photon timing approach was used, where a Gaussian approximation to
   the timing model likelihood was assumed. This does enable the estimation of
   TN properties for gamma-ray MSPs, but assumes a fixed pulse profile template.

   In this paper, we present a new method that extends the Gaussian process
   framework commonly used in radio pulsar timing to the gamma-ray data,
   allowing for efficient fitting even of complicated timing models, and robust
   estimation of stochastic noise processes, while accounting for uncertainty in
   the template pulse profile. We have implemented this method in a python
   package called \texttt{shoogle}\footnote{\url{https://github.com/coclar/shoogle}}\footnote{``Shoogle'' is a Scottish word that
   is both a noun and verb, meaning ``a gentle shake'' or ``to gently shake'',
   often to achieve some goal in a non-destructive manner. The pulsars we are
   timing here often have complicated ``shoogly'' timing residuals, that we wish
   to ``shoogle'' back into place.}, which uses the \texttt{PINT} pulsar timing
   package \citep{Luo2021+PINT,Susobhanan2021+PINT} to evaluate the pulsar
   timing model and its derivatives.

   This paper is organised as follows. Section~\ref{s:methods} describes our new
   method (with notation summarised in Appendix~\ref{a:glossary}): it begins
   with a summary of how the Gaussian process framework is used in pulsar timing
   (Section~\ref{s:gp_timing}); presents the gamma-ray timing problem
   (Section~\ref{s:gamma_ray_timing}) and our new method for tackling it
   (Section~\ref{s:gibbs_sampling}); and describes how this method can be used
   to time orbital period variations in spider binary systems
   (Section~\ref{s:opvs}), the problem that was our original motivation for
   developing this method. In Section~\ref{s:demonstrations} we test the method
   on both real and simulated pulsar data, demonstrating its accuracy. In
   Section~\ref{s:discussion} we discuss further possible uses for and
   extensions to this method, before briefly summarising in
   Section~\ref{s:conclusions}.

   \section{Methods}
   \label{s:methods}
   \subsection{Pulsar timing using Gaussian processes}
   \label{s:gp_timing}
   Pulsar timing analyses consist of finding a set of parameters $\vec{\rho}$ for a
   timing model $\Phi(t,\vec{\rho})$ that minimise the variance of the rotational
   phases for a set of observed pulse ToAs $\vec{t}$. Often, this consists of a
   constant spin-down model,
   \begin{equation}
     \Phi(t, \vec{\rho}) = \left(f\ t_{\rm psr}(t, \vec{\rho}) + \frac{1}{2} \dot{f}\, t_{\rm psr}(t, \vec{\rho})^2\right) \bmod{1}
   \end{equation}
   where $f$ is the pulsar spin frequency, $\dot{f}$ is the spin-down rate, and
   $t_{\rm psr}$ is the time at which a pulse detected at time $t$ was emitted
   by the pulsar (up to an arbitrary constant), taking into account various
   delays \citep{Edwards2006+Tempo2}: light travel time R\o{}mer delays due to
   the pulsar's location in a binary system, and due to Earth's motion around
   the solar-system; radio propagation delays; and relativistic delays due to
   light propagation through curved spacetime. All of these delays are unknown
   in advance, and must be parameterised and estimated from the data.

   Pulsar timing analyses usually make two approximations. First, the timing
   model is approximated as a linear expansion with parameter offsets
   $\vec{\theta} \equiv \vec{\rho} - \vec{\rho}_0$, relative to the initial model
   parameters $\vec{\rho}_0$,
   \begin{equation}
     \vec{\Phi}(\vec{t}, \vec{\rho}) \approx \vec{\phi}(\vec{t}, \vec{\theta}) \equiv
     \vec{\phi}_0(\vec{t}) - M(\vec{t})\,\, \vec{\theta}\,,
   \end{equation}
   where $\vec{\phi}_0(\vec{t}) \equiv \vec{\Phi}(\vec{t}, \vec{\rho}_0)$ are the
   pre-fit spin phases according to the initial model, and $M(\vec{t})$ is the
   \emph{design matrix}, containing the derivatives of the phase model at the
   observed arrival times, with respect to the parameters:
   \begin{equation}
     M_{ij}\,(\vec{t}) = \left.\frac{\partial \Phi(t_i, \vec{\rho})}{\partial \rho_j}\right|_{\vec{\rho} = \vec{\rho}_0}\,.
   \end{equation}
   This linear approximation is exact for some parameters, although not all
   \citep{Susobhanan2025+Vela}, but is generally a good approximation around the
   best-fitting models. For brevity, we shall drop the explicit dependences of
   $\vec{\phi}$ and $M$ on $\vec{t}$ and $\vec{\theta}$ hereafter.
      
   The second approximation that is nearly universally assumed is that the
   likelihood function for the data, $D$ (which consists of the ToAs, $\vec{t}$,
   and their uncertainties, $\vec{\sigma}$) is approximated as a multi-variate
   Gaussian, with log-likelihood:
   \begin{equation}
     \log p\!\left( D \mid \vec{\theta} \right) = -\frac{1}{2}\left[\log \left|2\pi C\right| + \left(\vec{\phi}_0 - M \vec{\theta}\right)^T C^{-1} \left(\vec{\phi}_0 - M \vec{\theta}\right)\right]\,,
     \label{e:gauss_like}
   \end{equation}
   where $C$ is the data covariance matrix. In the simplest case, this matrix contains $(f \vec{\sigma})^2$ along its diagonal, assuming all ToAs are independent of one another. However, it can also contain off-diagonal elements to account for instrumental effects that introduce correlations between observations.

   This Gaussian likelihood assumption is an excellent approximation for radio
   data, where the arrival time and a precise Gaussian uncertainty of a
   nominal pulse can be obtained by cross-correlating the (usually high S/N)
   folded data from an observation against a template pulse profile.

   With these simplifying assumptions, the timing model fit follows simple
   generalised least squares. The above log-likelihood can be re-arranged to
   \begin{align}
     \log p\!\left( D \mid \vec{\theta} \right) = -\frac{1}{2}\biggl[&\log\left|2 \pi C\right| + \vec{\phi}^T_0 C^{-1} \vec{\phi}_0 - \hat{\vec{\theta}}^T \Sigma^{-1} \hat{\vec{\theta}} \notag \\&+ \left(\vec{\theta} - \hat{\vec{\theta}}\right)^T \Sigma^{-1} \left(\vec{\theta} - \hat{\vec{\theta}}\right)\, \biggr]\,,
   \end{align}
   showing that this is also a multivariate Gaussian in $\vec{\theta}$, with
   maximum-likelihood estimator ${\hat{\vec{\theta}}}$ and covariance matrix $\Sigma$
   \begin{equation}
     \begin{aligned}
     \Sigma &= \left(M^T C^{-1} M\right)^{-1}\,,\\
     \hat{\vec{\theta}} &= \Sigma \, M^T C^{-1} \vec{\phi}_0\,.
     \end{aligned}
   \end{equation}

   Modern radio pulsar timing analyses have timing models with components that
   account for seemingly stochastic signals or sources of correlated noise:
   intrinsic red noise in the pulsar's spin rate \citep{Coles2011};
   time-dependent variations in the dispersion measure
   \citep{Larsen2024+DMGP,Iraci2024+DMGP}; variations in the density of the
   Solar wind \citep{Susarla2024+SWGP}; and the GWB
   \citep{vanHaasteren2009,Lentati2013+GWBGP}. Obtaining robust estimates of the
   properties of these processes can be crucial, either because the processes
   themselves are the scientific target (e.g. measuring the amplitude and
   spectrum of the GWB), or because these processes must be accounted for to
   obtain robust estimates for other timing properties of interest
   \citep{Coles2011}.

   This can be achieved by treating these effects as stochastic Gaussian
   processes, which have been extensively studied in the literature. An exact
   solution would be obtained by simply modifying the data covariance matrix,
   $C$, to reflect the expected correlations in the residuals due to these noise
   components. However, this can be expensive, as the cost of inverting $C$
   grows with the cube of the number of ToAs.

   Instead, in the pulsar timing framework, these Gaussian processes are
   normally fit using a low-rank approximation \citep{vanHaasteren2015}. This is
   done by including a set of basis vectors in $M$ for each noise component,
   with corresponding coefficients in $\vec{\theta}$ that have a Gaussian prior whose
   variance is determined by a set of hyper-parameters, $\vec{\lambda}$, quantifying these
   noise processes.

   In pulsar timing analyses, these basis vectors are usually chosen to be a
   Fourier-series expansion, in which case the hyperparameters describe a model
   power spectral density (PSD) that defines the prior variance on the Fourier
   amplitudes, although it has been shown recently that a sparse time-domain
   basis yields more accurate results \citep{Crisostomi2025+FFTFit}.

   For a Gaussian prior on $\vec{\theta}$, centered on $\vec{\theta}_0$ and with
   covariance matrix $\Lambda(\vec{\lambda})$, the posterior distribution for
   $\vec{\theta}$, 
   \begin{equation}
     \log p\!\left( \vec{\theta} \mid D, \vec{\lambda} \right) = \,\log p\!\left( D \mid \vec{\theta} \right) + \log p\!\left( \vec{\theta} \mid \vec{\lambda} \right) - \log p\!\left( D \mid \vec{\lambda} \right)\,,
   \end{equation}
   is
   \begin{equation}
     \begin{aligned}
     \log p\!\left( \vec{\theta} \mid D, \vec{\lambda} \right) = \,-\frac{1}{2}\biggl[&\log \left|2 \pi C\right| + \vec{\phi}^T_0 C^{-1} \vec{\phi}_0 - \hat{\vec{\theta}}^T \Sigma^{-1} \hat{\vec{\theta}}\\
         &+ \left(\vec{\theta} - \hat{\vec{\theta}}\right)^T \Sigma^{-1} \left(\vec{\theta} - \hat{\vec{\theta}}\right) \\
         &+ \log \left|2\pi \Lambda\right| + \left(\vec{\theta} - \vec{\theta}_0\right)^T \Lambda^{-1} \left(\vec{\theta} - \vec{\theta}_0\right)\, \biggr] \\
       &- \log p\!\left( D \mid \vec{\lambda} \right)\,. \\
     \end{aligned}
   \end{equation}
   Expanding the quadratics and completing the square in $\vec{\theta}$ shows that the posterior distribution on $\vec{\theta}$ given $\vec{\lambda}$ follows another multivariate Gaussian in $\vec{\theta}$:
   \begin{equation}
     \begin{aligned}
       \log p\!\left( \vec{\theta} \mid D, \vec{\lambda} \right) = -\frac{1}{2}\biggl[&\log\left|2\pi C\right| + \log\left|2\pi\Lambda\right| \\
         &+ \vec{\phi}^T_0 C^{-1} \vec{\phi}_0  + \vec{\theta}_0^T  \, \Lambda^{-1}  \, \vec{\theta}_0 - \bar{\vec{\theta}}^T  \, \Gamma^{-1}  \bar{\vec{\theta}} \\
         &+ \left(\vec{\theta} - \bar{\vec{\theta}}\right)^T\Gamma^{-1} \left(\vec{\theta} - \bar{\vec{\theta}}\right)\, \biggr] \\
       & - \log p\!\left( D \mid \vec{\lambda} \right)\,.
     \end{aligned}
   \end{equation}
   with the new covariance matrix and mean,
   \begin{equation}
     \begin{aligned}
     \Gamma &= \left(\Sigma^{-1} + \Lambda^{-1}\right)^{-1}\,,\\
     \bar{\vec{\theta}} &= \Gamma \left(\Sigma^{-1} \hat{\vec{\theta}} + \Lambda^{-1} \vec{\theta}_0\right)\,.
     \label{e:theta_bar}
     \end{aligned}
   \end{equation}

   Since this is another multivariate Gaussian, it is easy to analytically
   marginalise over $\vec{\theta}$ to obtain the marginal likelihood for $\vec{\lambda}$,
   \begin{equation}
     \begin{aligned}
       \log p\!\left( D \mid \vec{\lambda} \right) = -\frac{1}{2}\Bigl[&\log\left|2\pi C\right| + \log\left|2\pi \Lambda\right| - \log\left|2\pi \Gamma\right| \\
         &+ \vec{\phi}^T_0 C^{-1} \vec{\phi}_0 + \vec{\theta}_0^T  \, \Lambda^{-1}  \, \vec{\theta}_0 - \bar{\vec{\theta}}^T  \, \Gamma^{-1}  \, \bar{\vec{\theta}} \, \Bigr]\,.
     \end{aligned}
     \label{e:log_margL}
   \end{equation}

   In summary, these two approximations: that the model is linear in $\vec{\theta}$
   around an initial model, and that the ToAs have Gaussian uncertainties,
   combined with modelling noise effects as Gaussian processes, leads to a
   Gaussian posterior distribution on $\vec{\theta}$, which has the convenient
   property of having a mean, covariance matrix and marginal likelihood that can
   be computed analytically. The hyperparameters can therefore be
   estimated by sampling from the marginal likelihood
   $p\!\left( D \mid \vec{\lambda} \right)$.

   \subsection{Gamma-ray pulsar timing}
   \label{s:gamma_ray_timing}
   The second assumption, that the data points have individual Gaussian
   uncertainties, does not apply to gamma-ray pulsar data, and prevents the use
   of the methods outlined above.

   The gamma-ray data are discrete, consisting of the arrival times,
   estimated directions and energies of individual gamma-ray photons. Due to the
   limited angular resolution of gamma-ray telescopes, it is not possible to say
   with certainty which photons in the data set were emitted by the pulsar we
   are trying to time, or by an unrelated background source. Instead, we must
   include all photons whose estimated directions are consistent with the
   position of the target pulsar, within the energy-dependent uncertainties determined
   by the instrument response functions.

   While we cannot be certain, we can still estimate the probability that a
   given photon was emitted by the targeted pulsar, and can use this
   probability to weight the contribution of each photon to the timing
   analyses \citep{Bickel2008,Kerr2011}. These probability weights, $w$, are obtained
   from a model of the gamma-ray sky that predicts the photon flux, $G$, that
   the LAT will detect from each source (the target pulsar, other sources in the
   region, and the diffuse background models), as a function of photon energy
   and arrival direction, after accounting for the instrument response functions
   and the time-, direction- and energy-dependent exposure:
   \begin{equation}
     w_j = \frac{G_{\rm psr}\,(E_j, \hat{n}_j, t_j)}
     {G_{\rm psr}\,(E_j, \hat{n}_j, t_j) + \sum_{i} G_{i}\, (E_j, \hat{n}_j, t_j)}
     \label{e:weights}
   \end{equation}
   where, $E_j$, $\hat{n}_j$, and $t_j$ are the reconstructed energy, direction
   and arrival time of the $j$-th photon, respectively, and the sum in the
   second term of the denominator runs over all other sources in the model
   (other nearby point sources, diffuse interstellar emission, and isotropic
   background emission).

   The gamma-ray photons from a pulsar are not all emitted from the same
   rotational phase. Rather, the gamma-ray pulse profile, i.e. the distribution
   of the rotational phases at which each photon was emitted, often has multiple
   peaks, and typical duty cycles of tens of percent \citep{3PC}. The broad and
   multi-peaked pulse shapes mean that the likelihood function for the pulsar's
   rotational phase given the arrival time of a single photon is not a Gaussian,
   as we assumed for the case of radio ToA measurements, and the rotational
   phase is far more uncertain than the microsecond timestamping precision of a
   single photon arrival time \citep{Ajello2021}.

   The underlying pulse profile is also unknown \emph{a priori}. We must therefore
   assume a family of possible pulse profiles parameterised by $\vec{\tau}$, and
   either assume fixed values or marginalise over these, to find the posterior
   distribution on the timing model and hyperparameters,
   \begin{equation}
     p\!\left( \vec{\theta},\vec{\lambda} \mid D \right) = \int p\!\left( \vec{\theta},\vec{\lambda},\vec{\tau} \mid D \right) d\vec{\tau}\,.
   \end{equation}
   The data consists of photons from the target pulsar, and from other sources. The
   likelihood for this data set, given the assumed timing model and template
   pulse profile is therefore
   \begin{equation}
     p\!\left( D \mid \vec{\theta},\vec{\tau} \right) = \prod_j \left[ p\!\left( \phi_j \mid P, \vec{\tau} \right) p_j\left(P\right) + p\!\left( \phi_j \mid \bar{P} \right) p_j\left(\bar{P}\right)\right]
   \end{equation}
   where $P$ and $\bar{P}$ denote the conditions that the photon did or did not
   come from the pulsar, respectively, and $p_j\left(P\right)$ and
   $p_j\left(\bar{P}\right)$ are the prior probabilities of these conditions for
   the $j$-th photon in the data set. These prior probabilities are simply the
   probability weight, $p_j\left(P\right) = w_j$ calculated from Equation~\ref{e:weights},
   and its negation, $p_j\left(\bar{P}\right) = 1 - w_j$. The distribution of phases for
   photons that do not come from the pulsar is assumed to be uniform with $0
   \leq \phi < 1$ (i.e., unpulsed emission), $p\!\left( \phi_j \mid P \right) = 1$. The likelihood
   function is therefore \citep{Kerr2015}
   \begin{equation}
     p\!\left( D \mid \vec{\theta},\vec{\tau} \right) = \prod_j \left[ w_j \, p\!\left( \phi_j \mid P, \vec{\tau} \right) + (1 - w_j)\,\right].
     \label{e:likelihood}
   \end{equation}

   The distribution of photon emission phases from the pulsar
   $p\!\left( \phi_j \mid P, \vec{\tau} \right)$ is the template pulse profile defined by the
   parameters $\vec{\tau}$.  Typically this template is modelled as a mixture model
   of individual components, on top of a constant unpulsed emission level
   \citep[which is usually close to zero,][]{2PC}. In this work, we assume that
   the template pulse profile consists of $K$ symmetric wrapped Gaussian
   peaks. The template pulse profile parameters therefore consist of the
   amplitudes ($A$), phases ($\mu$) and widths ($\sigma$) of these peaks, and
   \begin{equation}
     \begin{aligned}
     p\!\left( \phi \mid P, \vec{\tau} \right) = &\sum_{k=1}^K \frac{A_k}{\sqrt{2\pi} \sigma_k} \sum_{\ell=-\infty}^{\infty} 
     \exp{\left(-\frac{1}{2}\frac{(\phi - \mu_k - \ell)^2}{\sigma_k^2}\right)} \\
     &+ \left(1 - \sum_{k=1}^K A_k\right)\,
     \end{aligned}
   \end{equation}
   where $\phi$ is measured in fractions of a rotation (rather than
   e.g. radians). In practice, we usually only need to sum over one or two wraps
   (i.e. $\ell \in \left\{-2,-1,0,1,2\right\}$).

   It is the non-Gaussian form of this likelihood function, a weighted sum of a
   constant term and a mixture-model template pulse profile, that means that
   Gaussian-process based radio pulsar timing procedures are not immediately
   applicable to gamma-ray data. The multi-modal likelihood function leads to
   a multi-modal posterior distribution, and while a significantly detected
   pulsar has an approximately Gaussian global maximum that is much higher than
   the noise floor of local maxima in the surrounding parameter space, the
   location and shape of this global maximum cannot be solved for
   analytically. This can be seen by taking derivatives of the logarithm of
   Equation~\ref{e:likelihood} with respect to $\vec{\theta}$: the Hessian matrix of
   second derivatives is not constant over $\vec{\theta}$, as it would be for a true
   multivariate Gaussian distribution.

   Existing methods \citep[e.g.][]{Pletsch2015+J2339} use the \texttt{emcee}
   algorithm \citep{emcee} to sample from $p\!\left(
   \vec{\theta},\vec{\lambda},\vec{\tau} \mid D \right)$.  However, this
   procedure has some drawbacks when noise models are required. First, when
   fitting for a very low-amplitude noise process (e.g. the stochastic GWB) one
   can run into the Neal's Funnel problem
   \citep{Neal2000+SliceSampling,Gundersen2025+NealsFunnel} when a
   hyperparameter describing the width of a distribution is poorly constrained
   by the data. For small widths, high log-likelihoods can be found within a
   very narrow region of the parameter volume, while lower log-likelihoods are
   found over a wider region for larger widths. After marginalising over the
   parameter volume, the marginal likelihood for the width hyperparameter should
   be flat over a wide range (since the data do not constrain this parameter),
   but this funnel shape can be very challenging for some sampling
   algorithms to deal with, without applying a careful decentering transform
   to the parameters.

   Second, noise components may require a very large number of basis vectors and
   corresponding parameters to accurately model, in which case the
   dimensionality of the timing model can also become problematic. Take for
   example the new gamma-ray redbacks found by
   \citet{Thongmeearkom2024+TRAPUMRBs}, with complicated orbital period
   variations. These require many tens of additional parameters to describe the
   unpredictable orbital phase evolution, rapidly increasing the dimensionality,
   and therefore slowing down numerical optimisation and sampling methods.

   In the photon-by-photon method used in \citetalias{GPTA}, this problem
   was tackled by assuming a fixed template pulse profile, evaluating the
   gradient and Hessian matrix of the Poisson log-likelihood with respect to the
   timing model parameters, performing a gradient-based optimisation and then
   approximating the likelihood function as a Gaussian.

   The Gibbs sampling approach described in the next section combines the
   benefits of both of these methods, allowing us to efficiently sample from the
   full posterior distribution for timing models with stochastic noise
   components with unknown hyperparameters, while additionally marginalising
   over the uncertain template pulse profile parameters.

   \subsection{Gibbs sampling}
   \label{s:gibbs_sampling}
   A common approach for dealing with mixture-model likelihoods like our
   gamma-ray pulsar timing likelihood is to introduce new \emph{latent
   variables} that assign individual data points to one or other component of
   the mixture model. After doing so, the conditional likelihood function given
   these assignments reduces to a product of more tractable
   distributions. \citet{Daemi2019+GPGMM} use this approach to extend Gaussian
   process techniques to deal with data that has a mixture model likelihood, and
   use the Expectation-Maximisation algorithm to obtain maximum-likelihood
   estimates for the hyperparameters. Inspired by this, we adopt the latent
   variable approach, but use the technique of Gibbs sampling to obtain samples
   from the posterior distribution of interest, rather than just a
   maximum-likelihood point estimate.

   To do this in our gamma-ray timing procedure, we introduce the vectors
   $\vec{z}$ and $\vec{m}$, each of which contains one element per photon. The
   elements in $\vec{z}$ can take one of $K+2$ values, where $K$ is the number
   of peak components in the template pulse profile. The remaining two values
   correspond to photons attributed to unpulsed flux from the pulsar, and those
   attributed to (unpulsed) unrelated fore/background sources. Given a reference
   epoch, every photon can be assigned to a specific rotation of the neutron
   star. The elements of $\vec{m}$ correspond to additional integer rotations
   (hereafter wraps) relative to the rotation count predicted by the initial
   timing model.

   The extended posterior distribution becomes
   \begin{equation}
     p\!\left( \vec{\theta}, \vec{\lambda}, \vec{\tau}, \vec{z}, \vec{m} \mid D \right)
     \propto p\!\left( D \mid \vec{\theta},\vec{\tau},\vec{z},\vec{m} \right)\,p\!\left( \vec{\theta} \mid \vec{\lambda} \right)\,
     p\!\left( \vec{z},\vec{m} \mid \vec{\tau} \right)\,p(\vec{\lambda})\,p(\vec{\tau})\,.
   \end{equation}

   At first glance, it appears as if we have just succeeded in making the
   process even more complicated! However, inspecting the conditional
   likelihood shows that this now takes on a
   much more convenient form:
   \begin{equation}
     \begin{aligned}
       p\!\left( D \mid \vec{\theta},\vec{\tau},\vec{z},\vec{m} \right)
       &= \prod_j \sum_{k=1}^{K} \frac{\delta_{z_j\, k}}{\sqrt{2\pi} \sigma_k} \sum_{\ell=-\infty}^{\infty} \delta_{m_j\,\ell} \exp\left(\frac{-\left(\phi_j - \mu_k - \ell\right)^2}{2\sigma_k^2}\right) \\
&= \prod_j \frac{1}{\sqrt{2\pi} \sigma_{z_j}} \exp \left(-\frac{\left(\phi_j - \mu_{z_j} - m_j\right)^2}{2\sigma_{z_j}^2}\right)
     \end{aligned}
   \end{equation}
   The convenience comes from the Kronecker deltas, $\delta_{z_j\, k}$, and
   $\delta_{m_j\, \ell}$ which select, for each photon, a single pulsar
   rotation, and single peak from the multi-modal pulse profile template, or
   assign that photon to the unpulsed background if $z_j > K$. This
   likelihood function is now a multi-variate Gaussian, which as we have seen in
   Section~\ref{s:gp_timing}, allows us to easily solve for the posterior mean
   and covariance, and compute the marginal likelihood for the hyperparameters.

   Unfortunately, while the elements of $\vec{z}$ and $\vec{m}$ each take on one of a
   small number of discrete values, we cannot feasibly marginalise over these
   analytically: a gamma-ray data set contains $N\approx$\,10,000 photons, and
   the number of possible vectors that $\vec{z}$ can take is $(K+2)^N$, an enormous
   number. Instead, we must find a way to sample over this joint posterior,
   obtaining representative samples of the subsets of $\vec{z}$ and $\vec{m}$ that are
   consistent with the data, which we can then numerically marginalise over.

   Fortunately, this latent-variable posterior distribution is amenable to the
   technique of Gibbs sampling \citep{Geman1984+Gibbs}. In this method, a set of
   samples from a complicated joint posterior distribution is built up by
   drawing samples from the simpler conditional distributions of individual
   parameters or subsets of parameters (block Gibbs sampling), conditioned on
   fixed values for the remaining parameters, and iterating over all
   parameters. This procedure produces a Markov Chain whose stationary
   distribution is the joint posterior distribution of interest.

   Instead of sampling from $p\!\left( \vec{\theta},\vec{\lambda},\vec{\tau},\vec{z},\vec{m} \mid D \right)$
   directly, we use block Gibbs sampling to build up samples from this
   distribution by iteratively sampling from the conditional distributions
   \begin{equation}
     p\!\left( \vec{\tau},\vec{z},\vec{m} \mid D,\vec{\theta},\vec{\lambda} \right)\,
   \end{equation}
   and
   \begin{equation}
     p\!\left( \vec{\theta},\vec{\lambda} \mid D,\vec{\tau},\vec{z},\vec{m} \right)\,.
   \end{equation}
   In the next subsections we derive these distributions, and show how we can
   easily draw samples from them.

   \subsubsection{Sampling the template pulse profile and latent parameters}
   Given new (or initial) values for $\vec{\lambda}$ and $\vec{\theta}$, we need
   to sample the template pulse profile parameters and latent variables from the
   conditional distribution\footnote{This distribution has no dependence on
   $\vec{\lambda}$, as the pulse profile shape depends only on the distribution
   of photon phases, which is solely determined by the previously-sampled value
   of $\vec{\theta}$} $p\!\left( \vec{\tau},\vec{z},\vec{m} \mid
   D,\vec{\theta}\right)$. To do so efficiently, we use a collapsed Gibbs
   sampling method \citep{vanHaasteren2014+Gibbs}, in which we factorise this
   distribution into
   \begin{equation}
     p\!\left( \vec{\tau},\vec{z},\vec{m} \mid D,\vec{\theta} \right) \propto p\!\left( \vec{\tau} \mid D, \vec{\theta} \right)\, p\!\left( \vec{z} \mid D, \vec{\tau}, \vec{\theta} \right)\, p\!\left( \vec{m} \mid D, \vec{\tau}, \vec{z}, \vec{\theta} \right),
     \label{e:tau_zm_collapsed}
   \end{equation}
   where the distribution $p\!\left( \vec{\tau} \mid D,\vec{\theta} \right)$ has been
   marginalised over the latent variables, and then draw $\vec{\tau}$, $\vec{z}$, and $\vec{m}$
   in order from the distributions on the RHS of
   Equation~\ref{e:tau_zm_collapsed}. Drawing $\vec{\tau}$ from this marginal
   distribution improves the mixing of the sampling chain, as it depends only
   weakly on the previous latent variable samples, through the timing model
   sample $\vec{\theta}$ that was chosen according to these samples, rather than
   having the parameters for each peak component being tightly constrained by
   the photons that have been assigned to that peak.

   The marginal distribution $p\!\left( \vec{\tau} \mid D,\vec{\theta} \right)$
   is simply the product of the original mixture model likelihood of
   Equation~\ref{e:likelihood} and the prior on $\vec{\tau}$. Since the number
   of template parameters is fairly small, and the likelihood is quick to
   evaluate, we can sample $\vec{\tau}$ from this distribution using the
   Metropolis-Hastings algorithm. This requires a well-chosen proposal
   distribution. We obtain this by running an initial tuning MCMC run using the
   \texttt{emcee} algorithm \citep{emcee} before starting the Gibbs sampling
   procedure. We choose the proposal distribution to be a multi-variate Gaussian
   distribution whose covariance matrix is that of the samples from this tuning
   MCMC run. We then run a MH-MCMC chain using this proposal distribution to
   estimate the autocorrelation time, providing an estimate for how many steps
   to take to obtain an approximately indepedent sample of $\vec{\tau}$ during
   the Gibbs sampling.

   After obtaining a sample of $\vec{\tau}$, we evaluate the likelihood
   $p\!\left( D \mid \vec{z},\vec{\tau},\vec{\theta} \right)$, which has been marginalised over $\vec{m}$. This  factorises over photons, with each term
   being simply the wrapped Gaussian likelihood of each component for the observed
   photon phase
   \begin{equation}
     p\!\left( D \mid z_j = k,\vec{\tau},\vec{\theta} \right) =
       \frac{1}{\sqrt{2\pi} \sigma_k}
       \sum_{\ell=-\infty}^{\infty} \exp{\left(-\frac{1}{2}\frac{(\phi_j - \mu_k + \ell)^2}
                                      {\sigma_k^2}\right) }
     \label{e:z_likelihood}
   \end{equation}
   if $k \leq K$ (for photons assigned to a pulsed component) or
   $p\!\left( D \mid z_j > K,\vec{\tau},\vec{\theta} \right) = 1$ otherwise (for photons assigned to
   the unpulsed component of the pulsar's emission, if $k=K+1$, or to the
   unpulsed background if $k=K+2$).
   The prior $p\!\left( \vec{z} \mid \vec{\tau} \right)$ is simply the relative amplitudes of
   the components, multiplied by the photon probability weight:
   \begin{equation}
     p\!\left( z_j = k \mid \vec{\tau} \right) = 
     \begin{cases}
       w_j\, A_k & {\rm if }\; k \leq K \,\\
       w_j\, (1 - \sum_k^K A_k) & {\rm if }\; k = K+1 \,\\
       (1 - w_j) & {\rm if }\; k = K+2\,.
     \end{cases}
     \label{e:z_prior}
   \end{equation}
   and so we can easily draw a value for each $z_j$ weighted by the product of
   these distributions.

   Finally, we carefully deal with photons whose phase $\phi$ lies close to the
   periodic boundary at $\phi = 0$ or $\phi = 1$ by drawing a sample of $\vec{m}$
   from the distribution
   \begin{equation}
     p\!\left( D \mid m_j=\ell, z_j=k, \vec{\tau}, \vec{\theta} \right) = \frac{1}{\sqrt{2\pi}\sigma_k}
     \exp{\left(-\frac{1}{2} \frac{\left(\phi_j - \mu_k - \ell\right)^2}{\sigma_k^2}\right)}\,.
     \label{e:wraps_likelihood}
   \end{equation}
   We can then define two vectors, $\vec{\nu}(\vec{z},\vec{m},\vec{\tau})$ and
   $\vec{\omega}(\vec{z},\vec{m},\vec{\tau})$, containing the central phases (including phase
   wraps) and widths of the Gaussian peaks that each photon has been randomly
   assigned to,
   \begin{equation}
     \begin{aligned}
       \nu_{j}(\vec{\tau},\vec{z},\vec{m}) &= \mu_{z_j} + m_j\,,\\
       \omega_{j}(\vec{\tau},\vec{z},\vec{m}) &= \sigma_{z_j}\,.
     \end{aligned}
   \end{equation}
   Photons that have been assigned to the background or the unpulsed part of the
   pulse profile can be considered to have $\omega_{j} = \infty$, although in
   practice we simply ignore these photons in the next sampling stages, as they
   now have no effect on the timing model.
   
   \subsubsection{Sampling the timing model and hyperparameters}
   Next, we need to draw samples of the timing model and hyperparameters from
   their joint posterior distribution, conditioned on a chosen template pulse
   profile and latent photon--component assignments. Again, we use the collapsed
   Gibbs sampling scheme, factorising the conditional distribution into
   \begin{equation}
     p\!\left( \vec{\theta},\vec{\lambda} \mid D,\vec{\tau},\vec{z},\vec{m} \right) \propto
     p\!\left( \vec{\lambda} \mid D,\vec{\tau},\vec{z},\vec{m} \right) \, p\!\left( \vec{\theta} \mid D,\vec{\lambda},\vec{\tau},\vec{z},\vec{m} \right)\,,
   \end{equation}
   where the (un-normalised) conditional distribution for $\vec{\theta}$ is
   \begin{equation}
     p\!\left( \vec{\theta} \mid D,\vec{\lambda},\vec{\tau},\vec{z},\vec{m} \right) \propto p\!\left( D \mid \vec{\theta},\vec{\tau},\vec{z},\vec{m} \right)\,p\!\left( \vec{\theta} \mid \vec{\lambda} \right)\,,
     \label{e:theta_posterior}
   \end{equation}
   and the marginal likelihood for $\vec{\lambda}$ is its integral,
   \begin{equation}
     p\!\left( D \mid \vec{\lambda},\vec{\tau},\vec{z},\vec{m} \right) = \int p\!\left( D \mid \vec{\theta},\vec{\tau},\vec{z},\vec{m} \right)\,p\!\left( \vec{\theta} \mid \vec{\lambda} \right) d\vec{\theta}\,.
   \end{equation}
   The likelihood function now takes on
   a much simpler form, a product of individual Gaussians:
   \begin{equation}
     p\!\left( D \mid \vec{\theta},\vec{\tau},\vec{z},\vec{m} \right) = \prod_j
     \frac{1}{\sqrt{2\pi}\omega_{j}} \exp{\left(-\frac{1}{2}\frac{(\phi_j -
         \nu_{j})^2}{\omega_{j}^2}\right)}
     \label{e:gibbs_likelihood}
   \end{equation}
   This is the same form as the Gaussian likelihood of
   Equation~\ref{e:gauss_like}, after replacing the data covariance matrix, $C$,
   with the diagonal matrix $\Omega$ whose elements are $\vec{\omega}^2$, and
   $\vec{\phi}_0$ with $\vec{\phi}_0 - \vec{\nu}$, to obtain a multi-variate Gaussian
   likelihood,
   \begin{equation}
     \begin{aligned}
       \log p\!\left( D \mid \vec{\theta},\vec{\tau},\vec{z},\vec{m} \right) = -\frac{1}{2}\Bigl[&\log \left|2\pi \Omega \right| \\
         &+ \left(\vec{\phi}_0 - \vec{\nu} - M \vec{\theta}\right)^T \Omega^{-1} \left(\vec{\phi}_0 - \vec{\nu} - M \vec{\theta}\right)\Bigr]\,,
     \end{aligned}
     \label{e:gibbs_likelihood_theta}
   \end{equation}

   Given the Gaussian prior $p\!\left( \vec{\theta} \mid \vec{\lambda} \right)$, the mean, covariance
   matrix and marginal likelihood $p\!\left( D \mid \vec{\lambda},\vec{\tau},\vec{z},\vec{m} \right)$ for the
   resulting posterior distribution on $\vec{\theta}$ can be then found analytically,
   using the results in Equations~\ref{e:theta_bar} and \ref{e:log_margL}. We
   first sample from $\vec{\lambda}$ from this marginal distribution, then sample
   $\vec{\theta}$ from the resulting conditional distribution. This order avoids the
   Neal's Funnel problem that could arise if directly sampling from
   $p\!\left( \vec{\theta},\vec{\lambda} \mid D,\vec{\tau},\vec{z},\vec{m} \right)$, which can be very large over
   very narrow ranges of $\vec{\theta}$.
 
   As we did with the template parameters, we perform a short pre-tuned MH-MCMC
   process in $\vec{\lambda}$, targeting the marginal likelihood
   $p\!\left( D \mid \vec{\lambda},\vec{\tau},\vec{z},\vec{m} \right)$ and with an appropriate prior
   $p(\vec{\lambda})$, and take the final point as our new sample for $\vec{\lambda}$.
   Finally, we draw values for $\vec{\theta}$ from the distribution of
   Equation~\ref{e:theta_posterior}, using our assumed value for $\vec{\lambda}$ from
   this MH-MCMC step. This sample of $\vec{\theta}$ defines a new realisation of the
   timing model $\vec{\phi}$.

   \subsubsection{Summary of Gibbs sampling procedure}
\begin{figure*}
     \centering
     \includegraphics[width=0.95\textwidth]{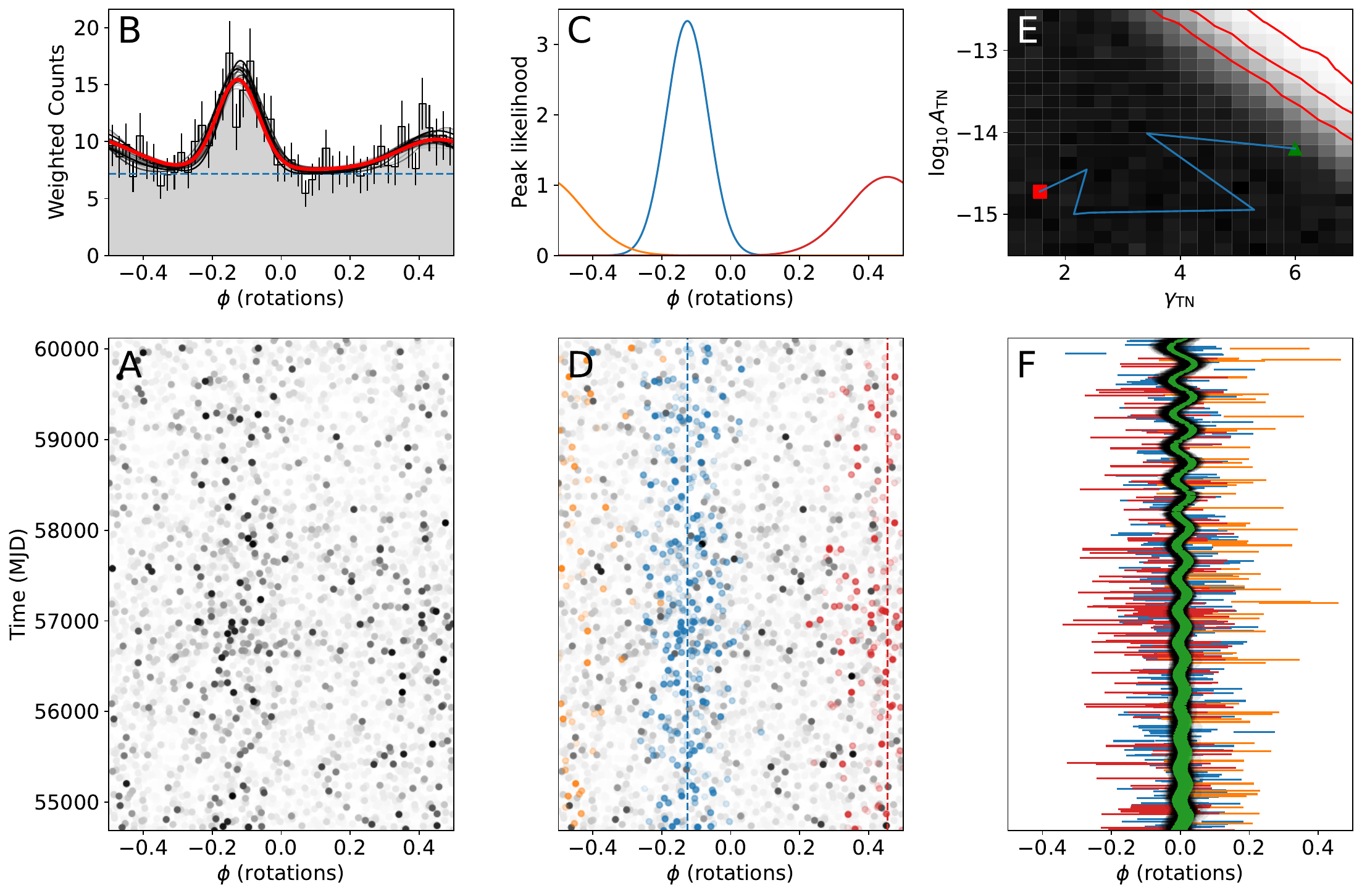}
     \caption{\label{f:gibbs_procedure} Illustration of the Gibbs sampling
       procedure applied to the binary MSP J1526$-$2744. \textbf{A:} photon
       phases are determined by the current timing model parameters, with
       probability weights indicated by the greyscale. \textbf{B:} a short
       Metropolis-Hastings MCMC chain is produced over the template pulse
       profiles (faint black curves), given these photon phases, and the final
       step is chosen as the next sample for $\vec{\tau}$ (solid red
       curve). \textbf{C:} The relative likelihoods as a function of phase for
       each Gaussian component in the template pulse profile are then used,
       alongside the photon weights, to determine the probabilities for
       assigning each photon to a single peak (or to the
       background). \textbf{D:} These assignments are performed randomly
       according to these probabilities, with the outcomes illustrated by the
       colours of photons in panel corresponding to the relevant peak in the
       panel above (or the grayscale for photons assigned to the
       background). \textbf{E:} A short MH-MCMC chain, starting from the red
       square and following the blue path to the green triangle, is run
       targeting the marginal likelihood for the hyperparameters, illustrated by
       the greyscale and red contour lines at the 1$\sigma$, 2$\sigma$ and
       3$\sigma$ levels. Here, a power-law timing noise model is assumed, with
       hyperparameters consisting of the log-amplitude $\log_{10}A_{\rm TN}$ and
       spectral index $\gamma_{\rm TN}$. The final sample in the chain is chosen
       as the next sample for $\vec{\lambda}$. \textbf{F:} A weighted
       least-squares fit is performed using Gaussian likelihoods for all photons
       assigned to peaks in the template, constrained by the prior distribution
       defined by the chosen sample of $\vec{\lambda}$. A single sample for the
       timing model (green curve) is chosen from the posterior distribution that
       results from this fit (illustrated by the faint black lines). Here, the
       sinusoidal curve whose amplitude grows with time indicates a significant
       detection of proper motion.  The photon phases in panel \textbf{A} are
       then updated by these resulting phase shifts, and the process repeats.  }
   \end{figure*}

   After drawing samples for $\vec{z}$,$\vec{m}$, $\vec{\tau}$, $\vec{\theta}$ and $\vec{\lambda}$ as above,
   one iteration of the Gibbs sampling is complete. To summarise, we repeatedly
   iterate through these Gibbs sampling steps, illustrated in
   Figure~\ref{f:gibbs_procedure}:

   \begin{enumerate}
     \item{Given the new (or initial) photon phases determined
   by $\vec{\theta}$, draw a new template pulse profile by running a short MH-MCMC chain
   over $\vec{\tau}$ with $p\!\left( \vec{\tau} \mid D,\vec{\theta} \right)$ as the target
   distribution, adopting the final sample.} \vspace{1ex}
     \item{Draw a random set of photon--component assignments $\vec{z}$ and phase wraps $\vec{m}$ from $p\!\left( \vec{z},\vec{m} \mid D,\vec{\tau},\vec{\theta} \right)$ using
       Equations~\ref{e:z_likelihood}, \ref{e:z_prior} and
       \ref{e:wraps_likelihood}.}\vspace{1ex}
     \item{Draw values
   for $\vec{\lambda}$, given $\vec{\tau}$, $\vec{z}$ and $\vec{m}$, by running another short MH-MCMC chain
   with $p\!\left( D \mid \vec{\lambda},\vec{\tau},\vec{z},\vec{m} \right) p(\vec{\lambda})$ as the target
   distribution (where the likelihood is computed using
   Equation~\ref{e:log_margL}).}\vspace{1ex}
     \item{ Given $\vec{\lambda}$, $\vec{\tau}$, $\vec{z}$ and $\vec{m}$, we draw
   a random sample for $\vec{\theta}$ from the multi-variate Gaussian distribution
   described by Equation~\ref{e:theta_bar}, and evaluate
   the timing model to find new photon phases $\vec{\phi}$.}
   \end{enumerate}

   After obtaining samples via this procedure, we evaluate the autocorrelation
   times for the sample chains of each element in $\vec{\tau}$, $\vec{\lambda}$, and
   $\vec{\theta}$, and iterate until a sufficient number of autocorrelation times
   (e.g. 1000) are achieved, to be confident that the sampling has converged on
   the target posterior distribution.

   \subsection{Modelling Orbital Period Variations}
   \label{s:opvs}
   The key benefit of the Gibbs sampling approach above is that it offers an
   efficient way to sample from high-dimensional timing models, and allows
   us to add and fit for stochastic noise components. This is because the latent
   variables turn the problem into a Gaussian process, from which we can easily
   draw samples of and marginalise over $\vec{\theta}$, no matter its dimensionality.

   One specific case where this efficiency is desirable, and in fact our
   original motivation for developing this method, is in timing spider binary
   systems, which contain MSPs with low-mass, semi-degenerate companion stars
   \citep{Roberts2013+Spiders}. These systems often exhibit significant
   long-term orbital period variations (OPVs). The origin of this behaviour,
   also seen in other types of close binary systems with white-dwarf or
   main-sequence primaries, is typically attributed to the Applegate mechanism,
   after \citet{Applegate1987} and \citet{Applegate1992}, in which gravitational
   quadrupole moment variations in the companion star caused by magnetic
   activity in the convective envelope couple with the orbital angular
   momentum. Variations like these are nearly universal among redback binary
   systems \citep[those with companion masses $M_{\rm c} \gtrsim
     0.1\,M_{\odot}$,
     e.g.][]{Deneva2016+J1048,Pletsch2015+J2339,Clark2021+J2039,Thongmeearkom2024+TRAPUMRBs,Corcoran2024+OPVGP},
   and are also seen in many black-widow binaries \citep[with $M_{\rm c}
     \lesssim 0.1\,M_{\odot}$,
     e.g.][]{Arzoumanian1994+B1957,Shaifullah2016+J2051,Voisin2020+J2051,BelmonteDiaz2025+J1544}. These
   OPVs significantly complicate gamma-ray timing efforts: their stochastic
   nature means that they cannot be predicted from a simple model, and can often
   require many tens of additional parameters in the timing model.
   
   This behaviour manifests in stochastic deviations $\Delta \Phi_{\rm orb}(t)$
   from the orbital phase $\Phi_{\rm orb}(t)$ predicted by a constant orbital
   frequency $f_{\rm orb}$, (possibly with a linear frequency change,
   $\dot{f}_{\rm orb}$, which can be caused by secular effects such as proper
   motion or acceleration in the Galactic potential),
   \begin{equation}
     \Phi_{\rm orb}(t) = f_{\rm orb}(t - T_{\rm asc}) + \frac{1}{2}\dot{f}_{\rm orb}(t - T_{\rm asc})^2 + \Delta \Phi_{\rm orb}(t)\,.
   \end{equation}
   where $t$ is the time measured at the binary barycentre, $T_{\rm asc}$
   is the time of the pulsar's ascending node, and $\Delta \Phi(t)$ is the
   accumulated orbital phase shift caused by the changes in the orbital
   frequency, $\Delta f_{\rm orb} (t)$\,
   \begin{equation}
     \Delta \Phi_{\rm orb}(t) = \int_{T_{\rm asc}}^t \Delta f_{\rm orb}(t^\prime)\, dt^\prime\,.
   \end{equation}

   To see what effect these variations have on the timing model, consider the
   first-order approximation to the R\o{}mer delay for a pulsar in a circular
   orbit, 
   \begin{equation}
     \Delta t_{\rm orb} \approx \frac{A_1}{c} \sin\left(2\pi \Phi_{\rm orb}(t)\right)\,.
   \end{equation}
   where $A_1$ is the pulsar's projected orbital semi-major axis. If we assume
   that $\Delta \Phi_{\rm orb}(t)$ is a small perturbation on top of the
   linearly-increasing orbital phase, then
   \begin{equation}
     \begin{aligned}
       \Delta t_{\rm orb} &\approx \frac{A_1}{c} \sin\left(2 \pi f_{\rm orb} (t - T_{\rm asc}) + 2\pi \Delta \Phi_{\rm orb}(t)\right) \\
       &\approx \frac{A_1}{c} \left[\sin\left(2 \pi f_{\rm orb} (t - T_{\rm asc})\right) + 2\pi \Delta \Phi_{\rm orb}(t) \cos\left(2 \pi f_{\rm orb}(t - T_{\rm asc})\right)\right]
     \end{aligned}
   \end{equation}
   where in the final line we have used the small-angle approximation. The
   variations in orbital phase caused by OPVs therefore produce sinusoidal
   delays at the orbital period, whose amplitude varies over time proportionally
   with $\Delta \Phi_{\rm orb}(t)$.

   Historically, OPVs have been modelled by a Taylor expansion of $f_{\rm orb}$
   around some reference epoch
   \citep[e.g.,][]{Ng2014+HTRU,Ridolfi2016+47Tuc,Pletsch2015+J2339} with the
   derivatives of the orbital frequency becoming the additional timing model
   parameters. However, this parameterisation suffers from strong correlations
   between parameters, and does not provide an obvious astrophysical
   interpretation.

   Starting from \citet{Clark2021+J2039}, we began instead modelling these
   variations as a Gaussian process in the orbital phase. We can then treat
   these variations as we do for TN, by adding appropriate basis
   vectors to the design matrix, and including additional hyperparameters that
   constrain the amplitudes of the coefficients of this basis expansion. In
   \citet{Clark2021+J2039}, we performed this fit in the time domain, with the
   hyperparameters directly parameterising the covariance matrix of $\Delta
   \Phi(t)$, using a sparse Gaussian process basis, and using the methods of
   \citet{Csato2002+SOGP} and \citet{Seiferth2017+GPMM} to fit the process
   despite the non-Gaussian likelihood.

   However, we found that this method broke down for more complicated pulsars,
   and often became stuck in local minima that lead to the timing model losing
   track of the pulsations and fitting random noise fluctuations instead. This
   problem was particularly pronounced for pulsars in wider orbits, due to the
   larger excess orbital R\o{}mer delay that orbital period variations cause.
   The original motivation for this Gibbs sampling method presented in this
   paper was to find a more reliable method for fitting this Gaussian process
   model.

   Here, we treat the OPVs in the same way that TN is usually treated
   in the pulsar timing literature, via a low-rank Fourier expansion with $F$
   frequencies,
   \begin{equation}
     \Delta \Phi_{\rm orb}(t, \vec{s}, \vec{c}) = \sum_{n=1}^{F} s_n \sin\left(\frac{2 \pi n (t - t_{\rm ref})}{T_{\rm obs}}\right) \
     + c_n \cos \left(\frac{2 \pi n (t - t_{\rm ref})}{T_{\rm obs}}\right)\,,
   \end{equation}
   where the $s_n$ and $c_n$ coefficients are the new timing model parameters
   added to $\vec{\theta}$, and $t_{\rm ref}$ is an arbitrary reference epoch.

   We can compute the corresponding basis vectors that must be added to the
   design matrix $M$ via the chain rule,
   \begin{equation}
     \begin{aligned}
     \left.\frac{\partial \vec{\phi}}{\partial s_n} \right|_{\vec{\theta} = \vec{0}} &=  \frac{\partial \vec{\Phi}_{\rm orb}}{\partial s_n} \left. \frac{\partial \vec{\phi}}{\partial \vec{\Phi}_{\rm orb}}\right|_{\vec{\theta} = \vec{0}}\,,\\
     &=  \sin \left(\frac{2 \pi n (\vec{t} - t_{\rm ref})}{T_{\rm obs}}\right) \left. \frac{\partial \vec{\phi}}{\partial \vec{\Phi}_{\rm orb}}\right|_{\vec{\theta} = \vec{0}}\,.
     \end{aligned}
   \end{equation}
   and similar for $c_n$ \textit{mutatis mutandis}. The remaining partial
   derivative on the RHS, $\partial \vec{\phi} / \partial \vec{\Phi}_{\rm orb}
   |_{\vec{\theta} = \vec{0}}$, is proportional to the derivative of the phase model with respect to $T_{\rm asc}$. We have implemented this Fourier basis for the orbital phase variations in
   the \texttt{PINT} pulsar timing software \citep{Luo2021+PINT,Susobhanan2021+PINT}, where it is named the ``ORBWAVES''
   model.

   As demonstrated by \citet{Bochenek2015}, the OPVs should, in principle, not
   correlate with TN, since the phase model derivative with respect to $T_{\rm
     asc}$ should be orthogonal to the pulse phase, provided there is sufficient
   coverage of orbital phases throughout the data.  This means that, if properly
   modelled, these orbital period variations should not introduce biases in TN
   measurements, or with a potential GWB signal, although they can reduce
   sensitivity to these effects by adding additional degrees of freedom to the
   modelling that can have the effect of smearing out the pulse profile.

   While this model is inherently non-linear, since the derivatives depend on
   the values of $\vec{S}$ and $\vec{C}$, we find in practice that the orbital
   phase deviations are so small, and can be constrained so well, that these
   derivatives can be assumed to be constant, provided the fitting is
   initialised at a point fairly close to the final model, although one or two
   iterations of fitting may be required to achieve this.

   This model is also non-physical. Changes in orbital period necessarily imply
   changes in the orbital semi-major axis, while a binary system whose companion
   star has a non-zero gravitational quadrupole moment will also necessarily
   have a small and rapidly-precessing eccentricity \citep{Voisin2020}. These
   effects are usually too small to be detected \citep[although
     see][]{Voisin2020+J2051}, particularly in gamma-ray data, and so we do not
   (yet) include them in this timing model, which only attempts to account for
   variations in the orbital phase.

   We also add additional hyperparameters to $\vec{\lambda}$ that place prior
   constraints on these Fourier coefficients. Specifically, $\vec{\lambda}$
   parameterises the PSD, $S_{\rm OPV}(f)$, of the variations in the pulsar's
   time of ascending node, which has units of time-squared per unit frequency,
   the same as the PSDs of TN processes. We adopt independent,
   uncorrelated Gaussian priors on the Fourier coefficients. The units for these
   coefficients are fractional orbits, so their prior variances are
   \begin{equation}
     \sigma^2_{s_n}(\vec{\lambda}) = \sigma^2_{c_n}(\vec{\lambda}) = \frac{f_{\rm orb}^2
       S_{\rm OPV}(n / T_{\rm obs}, \vec{\lambda})}{T_{\rm obs}}
   \end{equation}
   This diagonal covariance matrix prior is common in the pulsar timing
   literature, although \citet{AllenRomano2025} and
   \citet{Crisostomi2025+FFTFit} have shown that it is inaccurate, as it ignores
   correlations between different frequency bins due to spectral leakage from
   lower frequencies. For shallower spectra this problem is not severe, as the
   quadratic constant-frequency-derivative timing model absorbs low-frequency
   noise, but a sparse time-domain basis for the Gaussian process would likely
   recover noise parameters more accurately, particularly for steep spectra.

   \section{Demonstrations}
   \label{s:demonstrations}
   In this section, we demonstrate the Gibbs sampling timing method in three
   representative cases: (1) timing a ``simple'' binary millisecond pulsar
   without obvious TN or orbital period variations, showing that it
   returns posterior distributions that are consistent with those found with
   existing exact methods that sample the full posterior distribution; (2)
   simulations of millisecond pulsars with ``red'' TN processes,
   which show that our method correctly estimates the amplitudes and spectral
   indices of these noise processes; and (3) timing of the original
   black-widow binary pulsar, B1957$+$20, which exhibits significant orbital
   period variability, but which is one of the pulsars that contributes most
   strongly to the GPTA constraints on the GWB.

   \subsection{PSR~J1526$-$1744: basic timing model example}
   We first demonstrate the accuracy of the Gibbs sampling results by running it
   on a millisecond pulsar that does not require a complicated timing model, and
   comparing the results to those that were obtained using the existing
   \texttt{emcee}-based method that samples from the original posterior
   distribution, $p\!\left( \vec{\theta},\vec{\tau} \mid D \right)$.

   For this demonstration, we used the \textit{Fermi}-LAT data set used by
   \citet{Burgay2025} for the binary MSP J1526$-$2744.

   We here assume improper uniform priors on all timing model parameters, which
   consist of: right ascension $\alpha$, declination $\delta$ and corresponding
   proper motions, $\mu_{\alpha} \cos \delta$ and $\mu_{\delta}$; spin frequency
   $f$ and its first derivative $\dot{f}$; and orbital parameters $A_1$, $T_{\rm
     asc}$, $P_{\rm orb} = 1 / f_{\rm orb}$ and the two Laplace-Lagrangian
   parameters, $\eta$ and $\kappa$, parameterising eccentricity
   \citep{Edwards2006+Tempo2}. We do not include any noise components, and
   therefore have no hyperparameters to fit. We assume a uninformative Dirichlet
   distribution prior on the two template peak amplitudes, $A_k$, and unpulsed
   component ($1 - \sum_k A_k$), which ensures these sum to unity. We assume
   uniform priors on the peak centres, $p(\mu_k) = 1$, and a Jeffrey's prior for
   the peak widths, $p(\sigma_k) = 1/\sigma_k$. 

\begin{figure*}
     \centering
     \includegraphics[width=\textwidth]{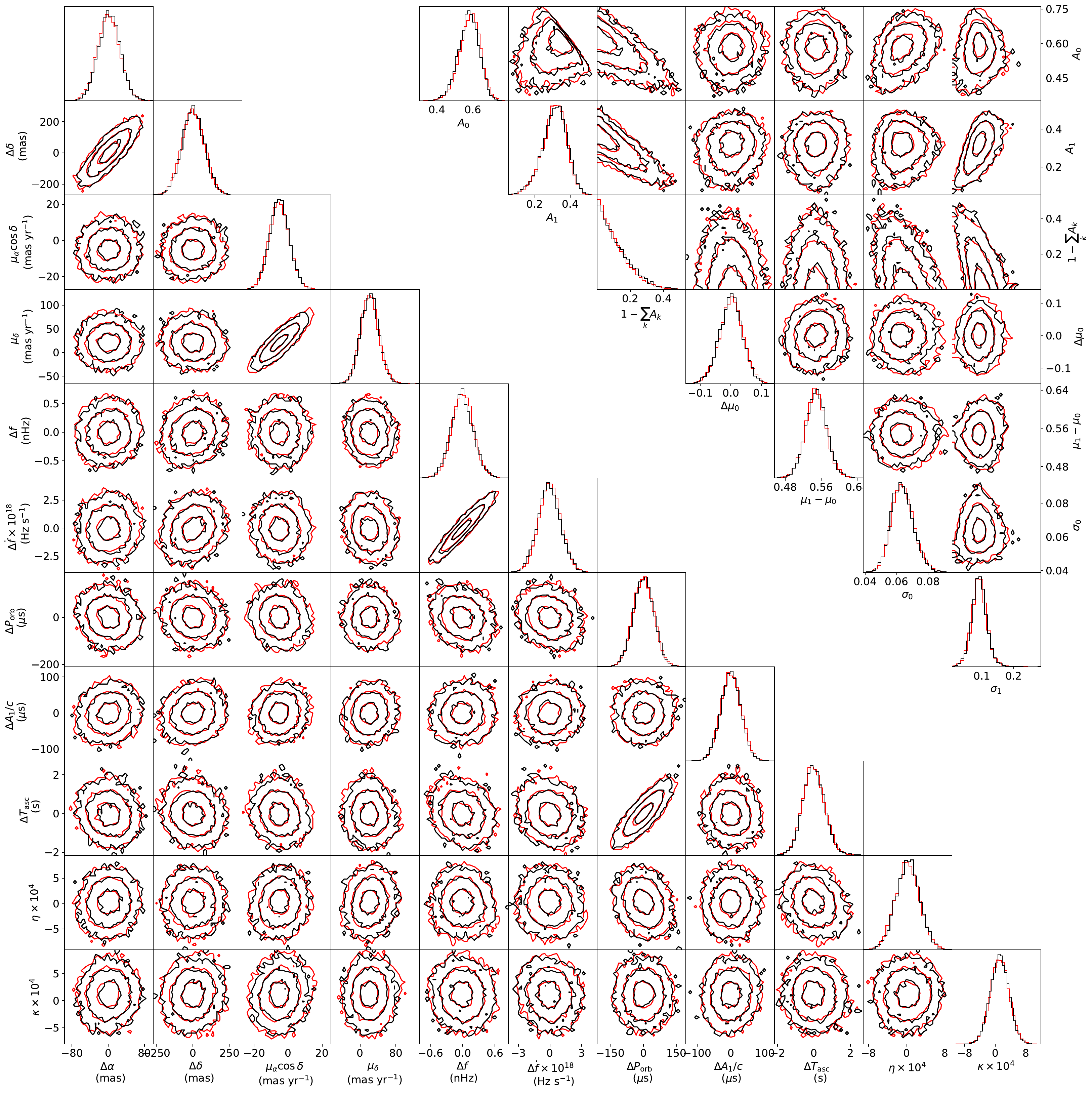}
     \caption{\label{f:J1526_comparison} Comparison between the posterior
       samples obtained via Gibbs sampling (red) and those obtained using the
       existing \texttt{emcee} method (black), for the binary millisecond pulsar
       PSR~J1526$-$2744. Template pulse profile parameters are shown in the
       upper right corner, while timing model parameters are shown in the lower
       left. There are no strong correlations between pairs of parameters across
       these two blocks. Contour lines are at the 1$\sigma$, 2$\sigma$ and
       3$\sigma$ level.}
   \end{figure*}

   The results are shown in Figure~\ref{f:J1526_comparison}. We find that the
   posterior distributions returned by the two sampling methods are
   indistinguishable from one another. This shows that the Gibbs sampling
   approach to marginalising over the latent variables successfully recovers the
   full posterior distribution.
   
   \subsection{Timing Noise Simulations}
   To test the robustness of the TN parameter estimates from the
   Gibbs sampling technique, we simulated and timed 200 pulsars with different
   levels of TN, and constructed Probability-Probability plots,
   \textit{PP-plots}, for the hyperparameters. A PP-plot is a common statistical
   visualization tool used to check for possible intrinsic biases in a recovery
   method for a specific parameter or set of parameters from a set of simulated
   datasets. The \textit{y}-axis corresponds to the fractional number of times
   the injected value lies within the estimated credible interval indicated on
   the \textit{x}-axis. When the recovery is unbiased, the PP-plot curve closely
   follows the diagonal. (See~\cite{ppplots2006, talts2020validating,
     wilkgnanadesikan68} for more details.)

   The simulation method implemented here takes as input the weights
   distribution, template pulse profile and timing model of a real pulsar and
   generates new photon arrival times. For the analysis presented here, the
   simulated pulsars have the same weights, timing model and template of PSR
   J1939+2134. For each simulation, we first randomize the order of the photon
   probability weights. For each photon, $j$, we then generate a uniform random
   number, $u_j$ between 0 and 1. If $u_j > w^{\prime}_j$, where $w^{\prime}_j$
   is the weight of the $j$-th photon in the randomised weights list, the photon
   is assigned to the background and its phase is drawn from a uniform
   distribution between 0 and 1. If $u_j < w_j$, the photon is assigned to the
   pulsar and the phase is sampled from the normalized pulse template. We now
   have a new weight and phase $\Phi_j^{\prime}$ for each photon. We compute the
   difference in phase between the new phase and the phase $\Phi_j$ of the
   original time of arrival, according to the chosen timing model, and update
   the time of arrival: $t_i^{\prime} = t_i + (\Phi_i^{\prime}-\Phi_i)/f_0$,
   where $f_0$ is the spin frequency. In this way, we obtained a set of
   realistic pulsar photon observations, but where the only noise present is due to
   the random sampling of the phase distribution.

    From here, we add additional time delays due to a TN process. Given a
    function that describes the power spectral density of the signal process as
    a function of frequency, we build the corresponding time-domain covariance
    matrix, and use that to define a zero-mean multivariate Gaussian
    distribution from which we sample a realization of time delays. To inject
    the TN in the simulated data, we simply add these time delays to the
    simulated noise-free photon arrival times.

    We simulated 200 pulsars with TN described by a smoothly broken powerlaw PSD:
    \begin{equation}
    \label{e:broken_pwl}
        S(f) = \frac{A^2}{12 \pi^2} \left(\frac{f_{\rm
            c}}{1~{\rm yr}^{-1}}\right)^{-\gamma} \left(1 +
        \left(\frac{f}{f_{\rm c}}\right)^2\right)^{-\gamma/2} \,{\rm yr}^3\,,
   \end{equation}
   with dimensionless amplitude $A$, cutoff frequency $f_{\rm c}$, and spectral index
   $\gamma$. The corresponding time-domain covariance function for this power
   spectrum is the Mat\'{e}rn function \citep{Rasmussen+GPML}.

   In our analysis, we simulated 12 years of observation and $f_c$ was set to an
   arbitrarily low value of $1 / (80\,{\rm yr})$, such that the noise process is
   effectively a pure power-law over the frequencies spanned by the data. For
   each realisation, we sampled the TN amplitude and slope from the uniform
   priors: $\log_{10}A_{\rm TN} \in [-16, -12]$ and $\gamma_{\rm TN} \in [1,
     7]$. TN was significantly detectable for amplitudes $\log_{10}(A_{\rm TN})
   \gtrsim -13$ at $\gamma \approx 4$, and at lower amplitudes for steeper
   spectra.
    
\begin{figure}
     \centering
     \includegraphics[width=1\columnwidth]{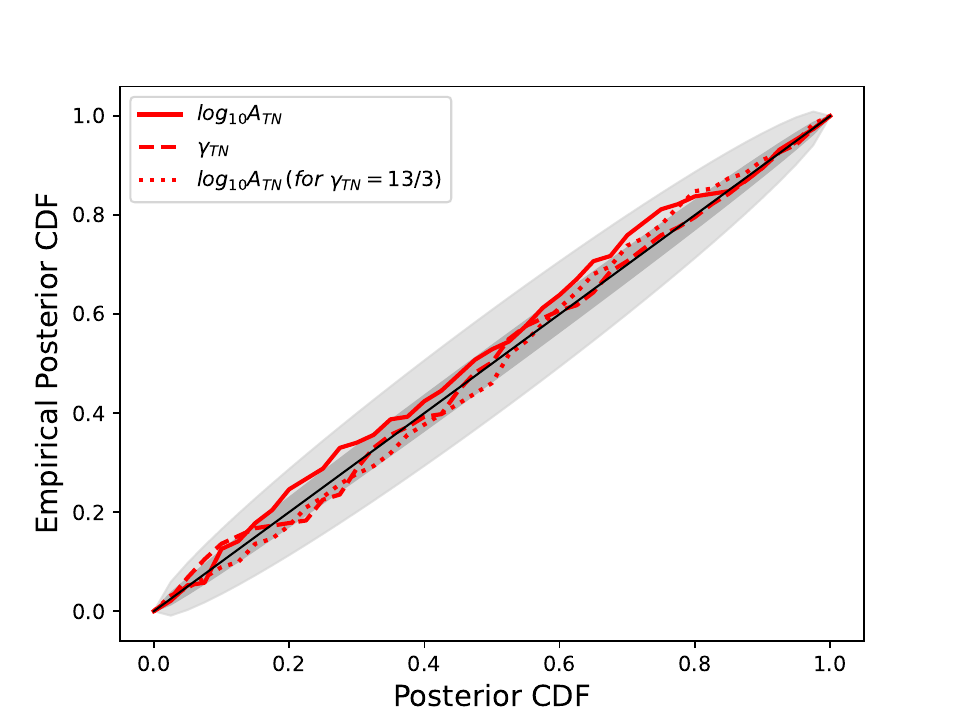}
     \caption{\label{f:PPplot} PP-plot for the recovery of the TN amplitude and
       slope, as defined in Eq.~\ref{e:broken_pwl}, from 200 simulated
       pulsars. We also show the case of the amplitude recovery from datasets
       where the slope was fixed to $13/3$ (predicted value for a GWB
       signal). The gray areas are the 1$\sigma$ and 3$\sigma$ confidence
       intervals from the predicted distribution (black diagonal).}
   \end{figure}

   In Figure~\ref{f:PPplot} we show the results for (i) the recovery of both the
   TN log-amplitude (\textit{solid} line) and slope (\textit{dashed} line) from
   a set of 200 simulated pulsars, and (ii) the TN log-amplitude recovery from
   another set of 200 pulsars whose TN slope was fixed to the simulated value of
   $13/3$, which is the expected value for a GWB produced by a population of
   supermassive black hole binaries (\textit{dotted} line). The results closely follow
   the diagonal of the plot and are always within the 3-$\sigma$ region of
   the expected distribution for an unbiased recovery. Thus, there is no
   indication for intrinsic biases in the recovery of those hyperparameters and
   their uncertainties.

   \subsection{PSR~B1957$+$20: Orbital period variations and timing noise modelling}
   \label{s:B1957}
   The previous subsections have demonstrated that the Gibbs sampling method
   correctly samples the full posterior distribution, and obtains reliable
   estimates for noise model hyperparameters in addition to the timing model.

   In this section we apply this method to one particular pulsar of interest,
   the original black-widow binary pulsar B1957$+$20 (a.k.a. PSR~J1959$+$2048).
   This pulsar is one of the fastest gamma-ray MSPs, and has the narrowest known
   gamma-ray pulse (in absolute terms) with a full-width at half-maximum of
   around 25$\mu$s \citep{Guillemot2012,Ajello2021}. This makes it one of the
   most important pulsars in the GPTA, with the third lowest upper
   limit on the GWB in the array.

   However, like many spider binaries, this binary also exhibits orbital period
   variations. In \citetalias{GPTA}, this pulsar was timed using a Taylor
   series expansion of the orbital frequency. It was found that fixing the
   orbital frequency derivatives to their nominal values reduced the upper limit
   on the TN/GWB amplitude by a factor of two, compared to a model
   where they were free to vary. Furthermore, this was one of only two pulsars,
   both black widows, for which excess white noise components were favoured in
   the model. These properties motivate more careful modelling of the OPVs and
   TN spectrum in this system.

   We performed the Gibbs sampling analysis on the data set used in
   \citetalias{GPTA}, fitting for both TN and OPVs using Fourier bases
   with constrained power spectra. We modelled both the OPV and GWB noise
   processes with broken power-law PSDs, as defined in
   Eq.~\ref{e:broken_pwl}. For the GWB, we fixed the spectral index to the
   predicted $\gamma_{\rm GWB} = 13/3$.  Since the cutoff frequency is expected
   to be lower than the lowest Fourier frequency for our data set, we fixed it
   to the arbitrarily low value of $f_{\rm c, GWB} = 10^{-2}$~yr$^{-1}$. To
   obtain an upper limit on the GWB amplitude, we placed a uniform prior on the
   amplitude (not the log-amplitude, which would lead to the upper limit
   depending on the lower bound of our prior), with $A_{\rm GWB} <
   10^{-10}$.

\begin{figure*}
     \centering
     \includegraphics[width=0.95\textwidth]{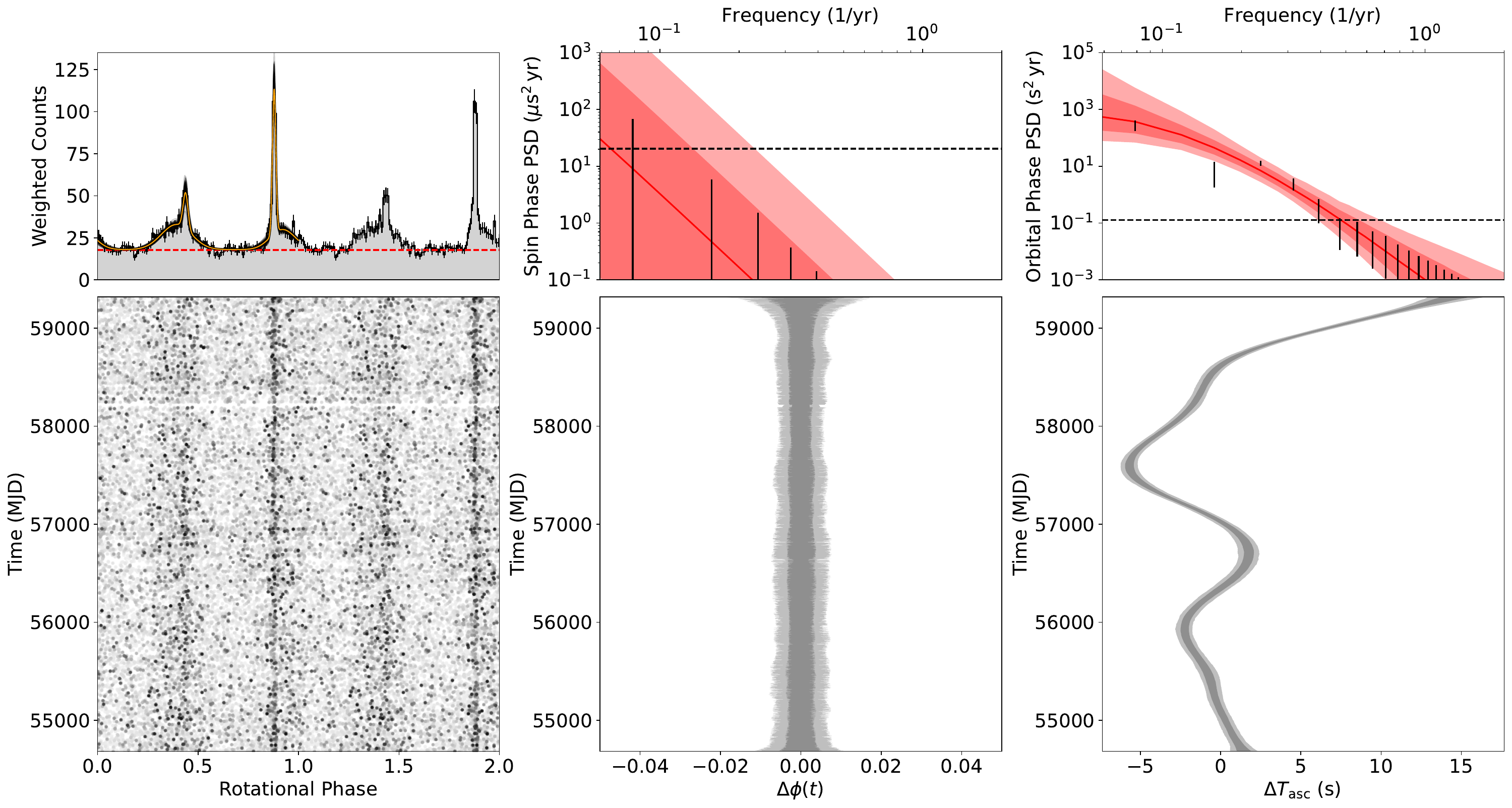}
     \caption{\label{f:B1957_demonstration} Results of Gibbs-sampling analysis
       on PSR~B1957$+$20. Left panels show the photon phases (bottom) and
       integrated pulse profile (top) according to the best-fitting timing
       model. Faint black curves on the pulse profile plot show individual
       posterior samples for the template pulse profile, with the best-fitting
       model shown in orange. Middle panels illustrate the posterior uncertainty
       on the photon phases (bottom), and on the PSD of the GWB-like timing
       noise component that we searched for (top). The dashed black line shows
       the estimated white-noise level, while the red shaded regions show the
       $1\sigma$ and $2\sigma$ constraints on a putative power-law noise
       component with spectral index $\gamma = 13/3$. Black lines show the
       posterior uncertainties on the individual Fourier powers - no significant
       power is detected above the white noise level at any frequency, and so
       these posteriors all closely follow the priors. Right panels show the
       same, but for orbital phase noise, in which a steep spectrum process is
       detected.
     }
   \end{figure*}

\begin{figure}
     \centering
     \includegraphics[width=\columnwidth]{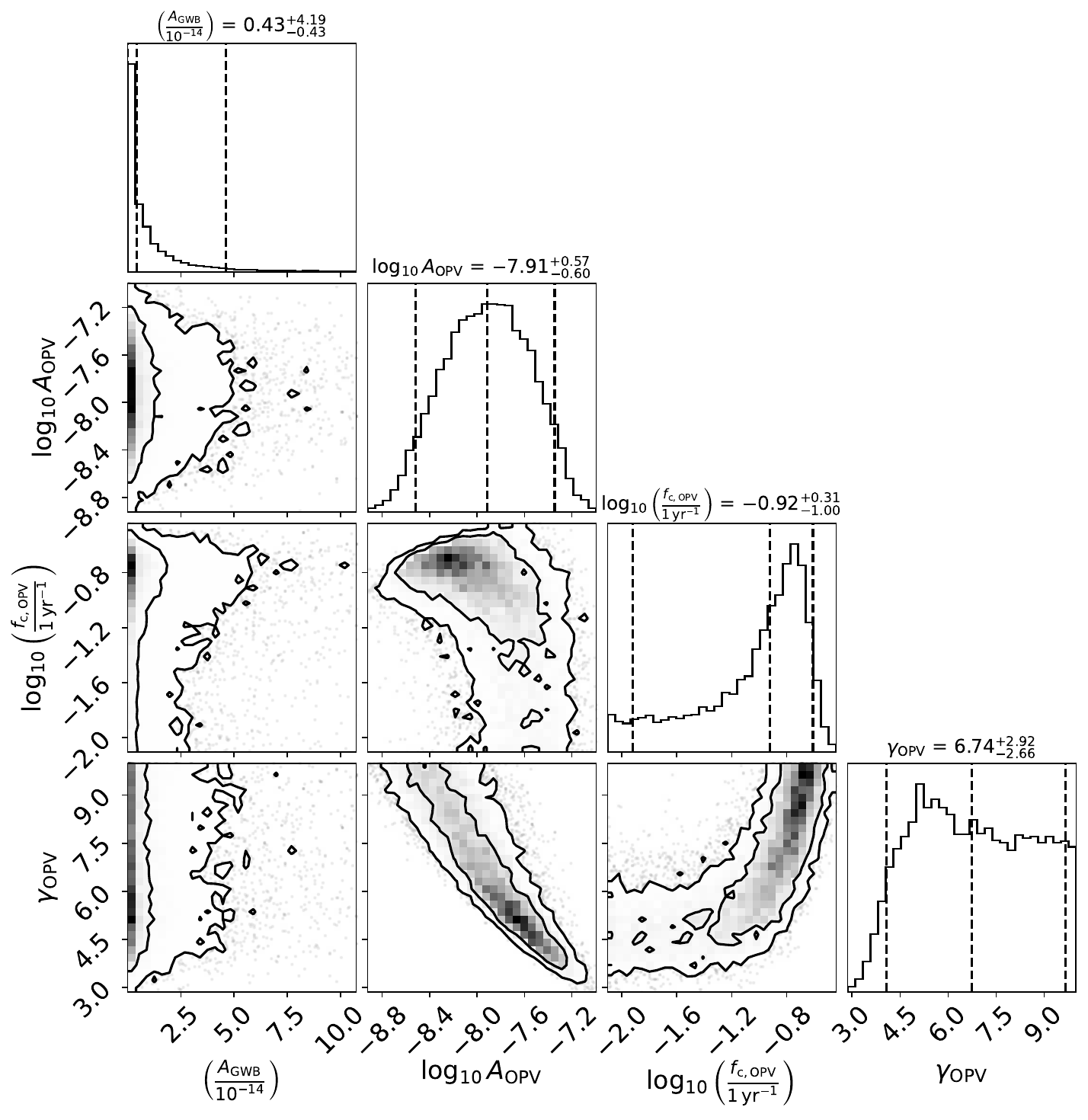}
     \caption{\label{f:B1957_corner} Posterior distribution for the GWB and OPV
       hyperparameters for PSR~B1957$+$20. Contours on the 2-D distributions are
       at the 1$\sigma$ and 2$\sigma$ levels, while dashed vertical lines on the 1-D
       marginal distributions indicate the 5\%, 50\% and 95\% quantiles.
     }
   \end{figure}

   For the OPV spectrum, we used a log-uniform prior on the amplitude $-15 <
   \log_{10} A_{\rm OPV} < -5$. We have found in other binaries
   \citep{Clark2021+J2039,Thongmeearkom2024+TRAPUMRBs} that a pure power-law
   model for the OPV spectrum does not always provide a good fit, and so we keep
   $f_{\rm c, OPV}$ free, with a log-uniform prior between $1 / {10 T_{\rm obs}}
   < f_{\rm c, OPV} < 10 / T_{\rm obs}$. For the spectral index, we adopted a
   uniform prior $0 < \gamma_{\rm OPV} < 10$.
   The results are illustrated in Figure~\ref{f:B1957_demonstration}, which
   shows the posterior distributions of the template pulse profile, the pulse
   and orbital phase variations and their PSDs, and in
   Figure~\ref{f:B1957_corner} which shows the posterior distributions on the
   GWB and OPV hyperparameters.  As in \citetalias{GPTA}, we find no evidence
   for a GWB-like TN component, with a 95\% confidence upper limit of
   $A_{\rm GWB} < 4.6 \times10^{-14}$.

   \section{Discussion}
   \label{s:discussion}

   We have presented a new method for timing gamma-ray pulsars, based on a
   latent variable approach for tackling the mixture-model likelihood function,
   and a Gibbs sampling scheme to jointly estimate posterior distributions for
   the timing model, noise parameters and the template pulse profile. In this
   section, we discuss possible applications and extensions of this method.

   One of the first broad applications of gamma-ray pulsar timing was the
   investigation of TN in young pulsars by \citet{Kerr2015}. This study used a
   ToA-based approach, which cannot always be applied to faint pulsars. Since
   then, the known gamma-ray pulsar population has grown, and the duration of
   \textit{Fermi}-LAT data has more than doubled. We used this Gibbs sampling
   method to time two young radio-quiet pulsars in \citet{Clark2025+EAH},
   finding that one had an unusual excess of TN at frequencies around
   1~yr$^{-1}$, which could be due to magnetospheric state switching like that
   seen in radio pulsars by \citet{Lyne2010} and \citet{Lower2025}.  A study of
   young pulsar TN over the longer data span, across the larger gamma-ray pulsar
   population, and using a more sensitive timing method may yield new insights.

   Another timely application is the GPTA's attempts to constrain
   the GWB. In \citetalias{GPTA}, two timing approaches were used to place limits on
   the amplitude of the GWB. The first used the same ToA-based approach as in
   \citet{Kerr2015}, with a cadence of a few months. The second method, which
   was found to be slightly more sensitive, particularly for fainter pulsars,
   used the full photon likelihood of Equation~\ref{e:likelihood}, making a
   Gaussian approximation to the posterior distribution on $\vec{\theta}$ that could
   be integrated to compute the marginal likelihood for $\vec{\lambda}$. This offers
   equivalent sensitivity to our Gibbs sampling method, and also enables noise
   parameter fitting, but does not contain a marginalisation over the uncertain
   template pulse profile. Furthermore, this method did not provide a way to
   search for correlated noise between pulsars following the predicted
   Hellings-Downs (HD) curve.

   \citet{Valtolina2024+FourierLike} have since developed a Fourier-domain
   likelihood for GWB-like inter-pulsar correlations. This method requires
   samples from the posterior distributions of the TN Fourier coefficients from
   single-pulsar analyses, with fixed noise hyperparameters, which can be
   obtained from our Gibbs sampler. Combining these two methods therefore allows
   us to obtain robust estimates for the parameters of an HD-correlated GWB
   signal, using an unbinned photon-by-photon approach where uncertainties in
   the template pulse profile and timing model parameters have been marginalised
   over. We explored this in Valtolina, S. et al. (2025b, PRD, submitted), where
   we use the Gibbs sampling scheme and Fourier-domain PTA likelihood of
   \citet{Valtolina2024+FourierLike} on the \citetalias{GPTA} dataset, finding a
   $\sim20\%$ increase in the upper limit for $A_{\rm GWB}$. While the updated
   constraint is slightly weaker, we found, using simulated gamma-ray PTA data
   sets produced using the same method applied here in
   Section~\ref{s:demonstrations}, that our Gibbs-sampling approach including
   the template pulse profile marginalisation provided a more robust estimate of
   the GWB amplitude, compared to the ToA-based approach which tended to
   underestimate it.

   A key motivation for the GPTA is that the gamma-ray data is insensitive to
   time-varying propagation effects that can bias radio ToA measurements
   (e.g. DM, scattering and solar-wind variations). Our Gibbs sampling method
   allows us to incorporate both radio and gamma-ray timing data simultaneously,
   by simply including the radio ToAs as additional Gaussian terms in
   Equation~\ref{e:gibbs_likelihood}, and adding appropriate basis vectors
   accounting for DM and associated noise processes to the design matrix (with
   radio-frequency dependent derivatives set to zero for all gamma-ray
   photons). This joint radio and gamma-ray timing capability has already been
   demonstrated for recently-discovered black-widow and redback binaries with a
   small number of radio ToAs \citep{Burgay2025,BelmonteDiaz2025+J1544}, but the
   full power of the method would be in jointly fitting radio and gamma-ray PTA
   data, where correlations between red noise and DM variations would be
   partially broken by the gamma-ray data.

   In Section~\ref{s:B1957}, we demonstrated the timing method on
   PSR~B1957$+$20, a black-widow binary with significant OPVs, which provides
   one of the most stringent gamma-ray constraints on the GWB. For this search
   we allowed all timing model parameters to vary in the fit and additionally
   marginalised over the template pulse profile, rather than assuming a fixed
   pulse shape. The resulting 95\% confidence upper limit of $A_{\rm GWB} < 4.6
   \times10^{-14}$ is larger than the \citetalias{GPTA} upper limit of $A_{\rm
     GWB} < 2.8\times10^{-14}$ that was obtained after fixing the astrometry and
   orbital parameters to their best-fitting values, but smaller than the $A_{\rm
     GWB} < 6.15 \times10^{-14}$ limit obtained with a similarly free timing
   model.

   This method also allows us to make quantitative statements about the OPV
   process. The OPVs here have a fractional amplitude of $\Delta P_{\rm
     orb}/P_{\rm orb} \sim 10^{-7}$, very similar to that found in earlier radio
   timing of this pulsar \citep{Arzoumanian1994+B1957}. Using the relations in
   \citet{Voisin2020+J2051}, we find that this fractional change in orbital
   period corresponds to a fractional change in the companion star's
   gravitational quadrupole moment of around $2\times10^{-5} k_2^{-1}$, where
   $k_2$ is the (unknown) apsidal motion constant. For the black-widow
   PSR~J2051$-$0827, \citet{Voisin2020+J2051} estimated a lower bound of $k_2
   \gtrsim 10^{-3}$. Assuming a similar value for PSR~B1957$+$20 implies
   fractional quadrupole moment changes of less than a few per cent. These
   values are very similar to those estimated for redback binaries
   \citep{Clark2021+J2039,Thongmeearkom2024+TRAPUMRBs}, despite the much lighter
   companion here.

   The cutoff frequency, $f_{\rm c,OPV}$ does not have a well-constrained lower
   limit, meaning that both pure power-law models (i.e. with $f_{\rm c,OPV} \ll
   1/T_{\rm obs}$), and those with cutoffs around $f_{\rm c,OPV} \approx 1 /
   (6~{\rm yr})$ are compatible with the data. If we restrict the posterior
   samples to those describing power-law like models (i.e., samples with $f_{\rm
     c,OPV} < 1 / T_{\rm obs}$) then we find $\gamma \approx 5 \pm 1$. We have
   found power-law like processes in other binaries with similar spectral
   indices \citep{Thongmeearkom2024+TRAPUMRBs,Burgay2025}, close to the value of
   $\gamma = 4$ that would result from a random walk in the orbital period or
   gravitational quadrupole moment. However, other binaries have OPV spectra
   with significant cutoffs and/or much steeper spectra
   \citep{Clark2021+J2039,Thongmeearkom2024+TRAPUMRBs,BelmonteDiaz2025+J1544}
   suggesting this property is not universal.

   \citet{Corcoran2024+OPVGP} and \citet{Rosenthal2025} timed six radio spider
   binaries in globular clusters, and used a similar Gaussian-process model to
   measure the OPV spectra for three of these, but found significantly shallower
   spectra, with $\gamma \approx 1$. We re-fit their orbital phase measurements
   for all six systems using a Gaussian process with a Matern covariance
   function (whose PSD is the smoothly broken power-law model used here), and
   found that all had $\gamma \gtrsim 4$, suggesting that the Lomb-Scargle
   method used by \citet{Corcoran2024+OPVGP} and \citet{Rosenthal2025} did not
   correctly account for spectral leakage from low frequencies
   \citep[e.g.][]{Coles2011}.

   So far, quantitive estimates of OPV processes such as these have mostly been
   applied to individual pulsars, but the \textit{Fermi} LAT has now detected
   gamma-ray pulsations from a population of more than 50 spider binaries
   \citep{3PC}. The Gaussian process model for OPVs, and the Gibbs sampling
   method presented here, have made timing studies of spiders in the
   \textit{Fermi}-LAT data tractable. A consistent treatment of this population
   may allow for the identification of trends between OPV parameters and other
   binary properties (e.g. orbital period, companion mass, etc.) that could
   provide valuable insight into the origin of this behaviour. Furthermore,
   these methods allow us to include several bright gamma-ray redback systems in
   the next GPTA data release, increasing its sensitivity and providing further
   independence from radio PTA efforts, where redbacks are usually excluded due
   to the detrimental effects of eclipses and their associated DM variations.

   \section{Conclusions}
   \label{s:conclusions}
We have presented a new method for timing gamma-ray pulsars in the
\textit{Fermi}-LAT data. This method is based on transforming the fitting
problem into a Gaussian process analysis through the introduction of latent
variables associating individual photons to components of a model pulse profile,
and using the technique of Gibbs sampling to marginalise over these latent
variables. This technique enables us to efficiently fit complicated timing
models and estimate parameters describing red TN and stochastic
OPVs, all while accounting for uncertainties arising from
the unknown intrinsic pulse profile shape. 

We have implemented this Gibbs sampling procedure in a python package called
\texttt{shoogle}, which is based on the \texttt{PINT} pulsar timing package. We
demonstrated that \texttt{shoogle} provides posterior distributions that are
consistent with those obtained using previous methods. We also simulated 200 new
realisations of the gamma-ray MSP J1939$+$2134 with additional TN, and
found that our method provides unbiased estimates for TN parameters
with robust uncertainties.

Finally, we demonstrated the new timing method on a 12-year data set for the
original black-widow binary PSR~B1957$+$20, where we were able to obtain an
upper limit on a red-spectrum GWB-like noise process of $A_{\rm GWB} <
4.6\times10^{-14}\,{\rm yr^{3/2}}$, similar to that obtained in
\citetalias{GPTA}, while simultaneously modelling the orbital period variations
as a Gaussian process. We found that the properties of these variations are very
similar to those observed in other redback and black-widow binaries.

Our new method was used alongside the work in \citet{Valtolina2024+FourierLike}
to re-analyse the \citetalias{GPTA} data set, to perform a search for correlated
noise produced by a stochastic gravitational wave background without having to
bin the LAT data into 6-month segments, resulting in a slightly higher, but
hopefully more robust, upper limit of $A_{\rm GWB} < 1.2\times10^{-14}\,{\rm
  yr^{3/2}}$ (Valtolina, S. et al., 2025, PRD, submitted). These methods will be used
to analyse the next GPTA data release, alongside additional improvements
such as incorporating photon energy-dependent pulse profile models to boost
sensitivity. Furthermore, our method provides a simple route for future joint
modelling of radio and gamma-ray pulsar observations, which may significantly
improve radio PTA sensitivity by partially breaking correlations between DM
variations and the GWB-induced residuals.

\begin{acknowledgements}
We are very grateful to Matthew Kerr for insightful discussions on gamma-ray
pulsar timing methods, and for helpful comments on the manuscript. We also thank
Xian Hou for reviewing the paper on behalf of the \textit{Fermi} LAT
Collaboration, and Kyle Corcoran for helpful discussion regarding their timing
of globular cluster redbacks.

The \textit{Fermi} LAT Collaboration acknowledges generous ongoing support
from a number of agencies and institutes that have supported both the
development and the operation of the LAT as well as scientific data analysis.
These include the National Aeronautics and Space Administration and the
Department of Energy in the United States, the Commissariat \`a l'Energie Atomique
and the Centre National de la Recherche Scientifique / Institut National de Physique
Nucl\'eaire et de Physique des Particules in France, the Agenzia Spaziale Italiana
and the Istituto Nazionale di Fisica Nucleare in Italy, the Ministry of Education,
Culture, Sports, Science and Technology (MEXT), High Energy Accelerator Research
Organization (KEK) and Japan Aerospace Exploration Agency (JAXA) in Japan, and
the K.~A.~Wallenberg Foundation, the Swedish Research Council and the
Swedish National Space Board in Sweden.

Additional support for science analysis during the operations phase is gratefully 
acknowledged from the Istituto Nazionale di Astrofisica in Italy and the Centre 
National d'\'Etudes Spatiales in France. This work performed in part under DOE 
Contract DE-AC02-76SF00515.
\end{acknowledgements}

\bibliographystyle{aa} 
\bibliography{ms} 

\begin{appendix}
  \label{a:glossary}
  \section{Glossary}
  \begin{table}[!h]
    \caption{Glossary of mathematical symbols used in this paper.
    \label{t:glossary}}
    \centering
    \begin{tabular}{lc}
      \hline
      \hline
      Description & Symbol \\
      \hline
      Times-of-arrival & $\vec{t}$ \\
      Timing model parameters & $\vec{\rho}$ \\
      Pre-fit timing model parameters & $\vec{\rho}_0$ \\
      Pre-fit rotational phases according to initial model & $\vec{\phi}_0$ \\
      Timing parameter offsets relative to initial model  & $\vec{\theta}$ \\
      Design matrix, $M_{ij} = \partial \phi(t_i,\vec{\rho}) / \partial \rho_j\,|_{\vec{\rho} = \vec{\rho}_0}$ & $M$ \\
      Rotational phases according to parameter offsets $\vec{\phi} = \vec{\phi}_0 - M \vec{\theta}$ & $\vec{\phi}$ \\
      Placeholder for all necessary data & $D$ \\
      Data covariance matrix & $C$ \\
      Timing parameter likelihood covariance matrix & $\Sigma$ \\
      Maximum likelihood timing parameters & $\vec{\hat{\theta}}$ \\
      Hyperparameters & $\vec{\lambda}$ \\
      Timing parameter prior covariance matrix & $\Lambda$ \\
      Timing parameter prior mean & $\vec{\theta}_0$ \\ 
      Timing parameter posterior covariance matrix & $\Gamma$ \\
      Timing parameter posterior mean & $\vec{\bar{\theta}}$ \\
      Number of Gaussian components in pulse profile & $K$ \\
      Pulse profile template parameters & $\vec{\tau}$ \\
      Amplitude of $k$-th Gaussian template component & $A_k$ \\
      Mean phase of $k$-th Gaussian template component & $\mu_k$ \\
      Width of $k$-th Gaussian template component & $\sigma_k$ \\ 
      Additional integer rotation (phase wrap) & $\ell$ \\
      Latent variables assigning photons to template components & $\vec{z}$ \\
      Latent variables assigning photons to rotations & $\vec{m}$ \\
      Means of per-photon Gaussian likelihood after latent variable assignments & $\vec{\nu}$ \\
      Variances of per-photon Gaussian likelihood after latent variable assignments & $\vec{\omega}$ \\
      Covariance matrix with $\vec{\omega}$ along diagonal & $\Omega$ \\
      Orbital phase & $\Phi_{\rm orb}$ \\
      Orbital frequency & $f_{\rm orb}$ \\
      Time of ascending node & $T_{\rm asc}$ \\
      Sine/cosine coefficients of Fourier series expansion of orbital phase variations & $\vec{s}, \vec{c}$ \\
      Power spectral density (PSD) of noise model & $S$ \\
      Amplitude of noise model at reference frequency of $1$\,yr$^{-1}$ & $A$ \\
      Corner frequency of smoothly broken power-law PSD model & $f_{\rm c}$ \\
      High-frequency spectral index of smoothly broken power-law PSD model & $\gamma$ \\
      \hline
    \end{tabular}
  \end{table}
\end{appendix}

\label{LastPage}
\end{document}